\documentstyle[aps,twocolumn]{revtex}

\begin{document}
\title{Anomalous Hopping Exponents of Ultrathin Films of Metals}
\author{N. Markovi\'{c},* C. Christiansen, D. E. Grupp,$^{\S }$ A. M. Mack,$^{\P }$
G. Martinez-Arizala, and A. M. Goldman}
\address{School of Physics and Astronomy, University of Minnesota, Minneapolis,\\
MN 55455, USA}
\date{December 18, 1999}
\maketitle

\begin{abstract}
The temperature dependence of the resistance $R(T)$ of ultrathin
quench-condensed films of Ag, Bi, Pb and Pd has been investigated. In the
most resistive films films, $R(T)=R_{0}\exp \left( T_{0}/T\right) ^{x}$,
where $x=0.75\pm 0.05$. Surprisingly, the exponent $x$ was found to be
constant for a wide range of $R_{0}$ and $T_{0}$ in all four materials,
possibly implying a consistent underlying conduction mechanism. The results
are discussed in terms of several different models of hopping conduction.
\end{abstract}




\section{INTRODUCTION}

In highly disordered materials, electrical conduction occurs by the hopping
of electrons between localized sites. This results in a thermally activated
electrical resistance of the form:

\begin{equation}
R(T)=R_{0}\exp \left( \frac{T_{0}}{T}\right) ^{x}  \label{RT}
\end{equation}
where $T$ is temperature, and $R_{0}$, $T_{0}$ and $x$ are constants which
depend on the disorder, the details of the interactions and the
dimensionality of the system. Simple activated hopping over a constant
barrier results in the Arrhenius form with $x=1$. For noninteracting
electrons, when the average hopping distance depends on temperature due to
the compromise between hopping to sites which are close in energy, but
farther away, Mott variable range hopping \cite{Mott} is expected, with $%
x=1/(d+1)$, where $d$ is the dimension. Efros and Shklovskii (ES) showed
that including Coulomb interactions between electrons results in a soft gap
in the density of states at the Fermi energy, which changes the variable
range hopping exponent to $x=1/2$ in all dimensions \cite{Efros}.

Hopping conduction has been investigated in a wide variety of materials,
such as doped semiconductors \cite{Dai,Zabrodski}, semiconducting
heterostructures \cite{VanKeuls}, amorphous metals \cite
{Lee,Shlimak,Liu,Hsu,Adkins}, magnetic materials \cite{Ioselevich} and
superconductors \cite{Dekker}. Both the Mott and the ES forms of variable
range hopping have been observed, as well as a crossover between the two
regimes \cite{VanKeuls,Lee}. It should be emphasized, however, that it is
often hard to distinguish between Mott and ES hopping, particularly in
experiments in which the resistance changes only by one or two orders of
magnitude. The unambiguous identification of the Mott or ES hopping can be
further complicated by factors which are usually neglected, such as the
granularity of the system, possible temperature dependence of $R_{0}$, or
correlations between electron hops.

While investigating the transport properties of ultrathin quench-condensed
films over the course of many years, we have often found that the resistance
of the thinnest films was thermally activated with $x\simeq 0.75$. A similar
hopping exponent has been reported by other authors \cite
{VanKeuls,Adkins,Gershenson,vanderPutten}, but has rarely been discussed 
{\it per se}. Since there is no theory which predicts this value of the
hopping exponent, its origin has been left an open question. Here we report
a detailed study of the temperature dependence of the resistance in very
disordered films of four different materials: Ag, Bi, Pb and Pd. The films
were grown in separate runs over several years, in different cryostats and
on different substrates, yet they all show the same thermally activated
resistance with an almost identical exponent. A careful analysis of the data
points to a new conduction mechanism in this regime, or perhaps calls for a
modification of the conventional picture. We compare our results with those
of other experiments and available theoretical calculations, and suggest
that the model that may possibly explain the anomalous hopping exponent is
the {\it collective} variable range hopping model of Fisher {\it et al}. 
\cite{Fisher}. Developed to describe vortices in superconductors, this
approach has not been considered before in the context of charge transport
in disordered electronic systems.

In Section II we survey the various models that have been considered in the
discussion of transport in disordered films. We also exhibit the mapping of
the model for collective vortex hopping onto the problem of charge transport
in disordered systems and estimate the value of the hopping exponent.
Experimental details of film growth and resistance measurements are given in
Sec. III. In Sec. IV, we analyze the temperature dependence of the film
resistance using several different methods to show that the exponent
obtained is really a property of the system, rather than a consequence of an
improper fit. The results are discussed and compared with other experimental
and theoretical work in Sec. V.

\section{SURVEY\ OF\ HOPPING\ MODELS}

In recent years, extensions of the basic variable range hopping model to
include percolation effects and correlations between electron hops have been
developed. Deutscher {\it et al}. \cite{Deutscher} proposed a hopping
mechanism which leads to a thermally activated resistance with $x$ close to $%
1/2$ without considering Coulomb interactions. The mechanism was based on
the superlocalization property of wavefunctions on incipient percolation
clusters \cite{Levy}, and may be relevant for atomically disordered systems
as well as for granular percolative structures. The detailed microstructure
of ultrathin quench-condensed films is still not known, and although these
films are usually considered to be homogeneous, they may actually contain
small grains or clusters. It is then possible that the electrons are
restricted to move on a sublattice which is fractal over some range of
length scales, and that their wavefunctions decay faster then exponentially
with distance. Based on this assumption, a hopping conductivity law has been
derived \cite{Deutscher} near the percolation threshold, which has the form
of Eq \ref{RT} with $x=3/7$. This is experimentally almost indistinguishable
from the Efros-Shklovskii law with $x=1/2$ if only the temperature
dependence of the resistance is studied, but can be identified through the
behavior of the parameter $T_{0}$ and the nature of the crossover to the
conventional Mott regime \cite{Deutscher}.

Generally speaking, there is no reason to assume that the prefactor $R_{0}$
in Eq. \ref{RT} is independent of temperature. Van Keuls {\it et al}. \cite
{VanKeuls} studied the resistivity in a gated $GaAs/Al_{x}Ga_{1-x}As$
heterostructures as a function of temperature, electron density and magnetic
field. Assuming the prefactor of the form:

\begin{equation}
R_{0}=bT^{m}  \label{R0}
\end{equation}
where $b$ and $m$ are constants, these workers fit Eq. \ref{RT} to their
data with $x=1/3$ in low magnetic fields and with $x=1/2$ in high magnetic
fields. The same crossover was observed as a function of electron density
and temperature, and it scaled with the separation between the electron
layer and the nearby screening gate, as predicted by Aleiner and Shklovskii 
\cite{Aleiner}. In addition to introducing a temperature dependent
prefactor, this experiment also raised the issue of the importance of
correlations between the electron hops.

In an electron glass, where the screening length is long \cite{Monroe} and
the interactions long-ranged, electron hopping may be correlated \cite
{Pollak}. Excitations can leave the system far from equilibrium \cite
{Ovadyahu} and relaxation occurs through the rearrangement of charge. The
energy of a single electron hop may then be significantly reduced by the
motion of the surrounding charges. At sufficiently low temperatures, such
collective hopping might be the dominant conduction mechanism.

In the analysis of their data, Van Keuls {\it et al}. \cite{VanKeuls}
assumed that the number of configurations of occupied states reached by the
correlated hopping of a number of electrons is proportional to the
single-particle density of states. In that case, the qualitative behavior of
the resistance remains unchanged, and the effects of correlations enter
through the constants which determine the parameter $T_{0}$ in different
regimes.

Yet another issue which can be relevant in extremely thin films is the
possibility that the electrons might interact logarithmically rather than as
1/r. As shown by Keldysh \cite{Keldysh}, the range of the logarithmic
interaction is given by:

\begin{equation}
r_{\log }=\frac{\kappa }{\kappa _{s}+1}d  \label{r}
\end{equation}
where $\kappa $ and $\kappa _{s}$ are the dielectric constants of the film
and the substrate, respectively, and $d$ is the film thickness. One possible
consequence of the logarithmic electron-electron interactions is that the
Coulomb gap in the density of states (linear in 2D) may change to an
exponential form \cite{Shklovskii}. This leads to a modified variable range
hopping law with a temperature dependent exponent.

Alternatively, the behavior of the logarithmically interacting electrons
might be similar to that of vortices in 2D superconductors, which are known
to interact logarithmically. Collective variable range hopping of vortices
in disordered thin-film superconductors was studied by Fisher {\it et al}. 
\cite{Fisher}. Including the effect of correlations, these authors found
that multivortex hopping results in a lower energy than single vortex
hopping. They also suggested that such multiparticle hopping might dominate
single particle hopping even in the case of inverse power law interactions.

This approach may then be mapped onto a disordered two-dimensional system of
charges. Following the arguments of Fisher {\it et al}. \cite{Fisher}, the
energy $U(r)$ of a multi-particle excitation of length $r$, can be estimated
to be:

\begin{equation}
U(r)\approx K(\frac{l}{r})^{1/2}  \label{Ur}
\end{equation}
where $l$ is the distance between charges, and $K$ is the bare
single-particle excitation energy. The latter is equal to $e^{2}/\kappa a$,
where $a$ is the localization length and $\kappa $ is the dielectric
constant. The simultaneous hopping of many charges may then result in a
lower energy than the hop of a single charge.

The electrical resistance is a product of the probability for an electron to
tunnel a distance $r$, $\Gamma _{r}$, and the probability for an excitation
with the energy $U(r)$ to occur, $\Gamma _{U}$:

\begin{equation}
R\propto 1/\Gamma _{r}\Gamma _{U}  \label{R}
\end{equation}

A lower bound for the multihop rate can be estimated as follows: if all of
the $(r/l)^{2}$ electrons in the $r$ by $r$ region hop a distance comparable
to the spacing $l$, then the rate should be proportional to the single-hop
rate , $exp(-l/a)$, raised to the power of the number of electrons,
resulting in:

\begin{equation}
\Gamma _{r}\propto e^{-(l/a)(r/l)^{2}}  \label{gammar}
\end{equation}

The probability for an excitation of energy $U(r)$ is proportional to $%
exp(-U(r)/T)$, or using Eq. \ref{Ur}:

\begin{equation}
\Gamma _{U}\propto e^{-(K/T)(l/r)^{1/2}}  \label{gammaU}
\end{equation}

The minimum resistance is obtained when the hopping distance $r_{hop}$ is:

\begin{equation}
r_{hop}^{2}\approx al(\frac{T_{0}}{T})^{4/5}  \label{rhop}
\end{equation}
where

\begin{equation}
T_{0}=K(l/a)^{1/4}  \label{T0}
\end{equation}

Substituting Eqs. \ref{gammar} through \ref{T0} back into Eq. \ref{RT}
results in a hopping form such as that of Eq. \ref{RT} with $x=4/5$.

In the other limit, the minimum number of electrons participating in a
collective hop could be taken as $(r/d)$, which leads to $x=2/3$. The
collective variable range hopping mechanism may therefore result in a
resistance of the form of Eq. \ref{RT}, where the range of exponents is $%
2/3<x<4/5$, depending on the fraction of electrons participating in the
process.

\section{EXPERIMENTAL METHODS}

The temperature dependence of the resistance has been studied in ultrathin
quench-condensed films of Ag, Bi, Pd \cite{Liu} and Pb \cite
{Martinez1,Martinez2}. The films were deposited on liquid helium cooled
substrates and resistance measurements were performed {\it in situ} at
temperatures down to $0.15K$. Ultra-high vacuum conditions and temperatures
below $20K$ were sustained throughout each run, in order to avoid
contamination or crystallization. The substrates were glazed alumina (for Bi
and Pd films) or $SrTiO_{3}$ (100) (for Ag and Pb films). The $SrTiO_{3}$
(100) substrates were $0.75mm$ thick and had a $100nm$ thick Au gate on the
back. Such a gate can be used to study the response of the film to a
perpendicular electric field and was used to establish glass-like behavior
in the most resistive films \cite{Martinez1,Martinez2}.

Films were deposited in thickness increments between $0.05$ and $0.5\AA $ on
top of a thin germanium layer $(5-10\AA )$. (The Pd films were the exception
to this as they were deposited directly onto glazed alumina substrates where
they became connected at monolayer coverage.) Films grown on amorphous Ge
are believed to be homogeneous, since they become connected at an average
thickness of about one monolayer \cite{Strongin}. The thicknesses of the
films studied ranged from $5\AA $\ up to $15\AA $. These nominal values of
the film thickness are calculated from the deposition rate, which was
measured using a quartz crystal monitor placed in the vicinity of the
substrate. The first low-temperature scanning tunneling microscopy studies
of the morphology of films grown in a similar manner indicate that the
thinnest films may indeed be homogenous, while the thicker ones contain
small clusters \cite{Ekinci}.

Resistance measurements were carried out using a standard dc four-probe
technique. Very low bias currents $(<10nA)$ were used to avoid Joule heating
of the sample and to make sure that the voltage across the sample was a
linear function of the applied current. When measuring very resistive films (%
$10^{4}-10^{8}$ $\Omega )$, because of the long time constants of the
circuit, it was necessary to monitor the voltage as a function of time after
the current was changed, and allow adequate time for the voltage to
stabilize. To avoid the voltage offset errors due to thermal EMFs, both
polarities of the current were used to determine the resistance.

The resistance of a series of Ag films was also studied in a magnetic field.
Magnetic fields of up to 20kG (12 kG) were applied in direction parallel
(perpendicular) to the plane of the substrate using a split-coil
superconducting magnet. In a regime where the anomalous hopping exponent is
observed, the resistance was found to be independent of magnetic field.

\section{RESULTS AND ANALYSIS}

The temperature dependence of the sheet resistance (resistance per square)
for series of Ag, Bi, Pb and Pd films is shown in Fig. \ref{logR}. The
logarithm of the sheet resistance, plotted as a function of $T^{-0.75}$
follows a straight line for each film, indicating that the resistance is
thermally activated with $x\approx 0.75$. Using other values for $x$, such
as $1$, $1/2$ or $1/3$ yielded considerably larger deviations from the data.

Since the prefactor in Eq. \ref{RT} is generally expected to be
temperature-dependent $(m\neq 0$ in Eq. \ref{R0}$)$, we attempted to fit the
data using different combinations of $m$ and $x$. As shown in Fig. \ref{xm},
using values of $m$ greater than zero actually increased the error of the
fit. The maximum deviation in the fit of the combinations of Eqs. 1 and 2 to
the data became much larger than the noise in $R$ as $m$ was increased.
Furthermore if values of $m$ were chosen to force either Mott or ES hopping
exponents of $x=1/3$ or $1/2$, respectively, the quality of fits as measured
by chi squared would be significantly worse than that with $m=0$, in
contrast with the findings of Van Keuls {\it et al}. \cite{VanKeuls}.

Assuming that the hopping exponent is $x\approx 0.75$, the activation energy 
$T_{0}$ can be extracted from the fit to Eq. \ref{RT}. The values of the
parameter $T_{0}$ for different films of all four materials are shown in
Fig. \ref{T0vsR} as a function of $R_{14K}$, which is the sheet resistance
measured at $14K$. This quantity is inversely proportional to the film
thickness, so by using $R_{14K}$ instead of the thickness, one can avoid
systematic errors in the nominal thicknesses of the films of different
materials. It is apparent in Fig. \ref{T0vsR} that $T_{0}$ changes greatly
as $R_{14K}$ (and therefore also the film thickness) changes, ranging from
around $100K$ for the thinnest films, to around $10K$ for the thickest
films. The same qualitative and quantitative behavior was found for all four
materials.

A more direct method of determining the exponent $x$ (which is exact under
the condition $m<<\left( T_{0}/T\right) ^{x}$) has been developed by
Zabrodskii and Zinov'eva \cite{Zabrodski}. The method is based on defining
the function $w=-d(\log R)/d(\log T)$. If $R$ is given by Eq. \ref{RT}, then 
$\log w\propto -x\log T$. By plotting $\log w$ as a function of $\log T$, $x$
can be easily extracted from the slope of the resulting straight line. The
benefit of the Zabrodskii-Zinov'eva approach is the simplicity of fitting a
line rather than a complicated function with up to four adjustable
parameters($b,$ $m$, $T_{0\text{ }}$and $x$). Once it has been determined
that $m=0$, this method eliminates the danger of finding a local minimum
instead of the best fit. The results of determining $x$ this way are shown
in Fig. \ref{zab} . For very resistive films, plotting $\log w$ vs. $\log T$
indeed yielded straight lines. The values of $x$ varied slightly between
different materials, from $x\approx 0.7$ for Ag to $x\approx 0.8$ for Bi.
Remarkably, the value of $x$ did not change between different films in the
same series over a significant range of sheet resistances, as shown in Fig. 
\ref{xR}.

Even though $R_{14K}$ and $T_{0}$ change from film to film, as more material
is added to increase the average thickness and decrease the sheet resistance
of the film, $x$ stays constant over three orders of magnitude in $R_{14K}$.
For thicker films (smaller $R_{14K}$), $x$ drops rather abruptly to a value
between $1/3$ and $1/2$. In this regime, the Zabrodskii-Zinov'eva plots no
longer produce straight lines, indicating that the hopping exponent changes
as a function of temperature. Further increase of the film thickness leads
to another crossover to a weakly localized regime where the temperature
dependence of the resistance is logarithmic (not shown in Fig. \ref{xR}).

\section{DISCUSSION}

An activated temperature dependence of the resistance with an anomalous
hopping exponent $x\approx 0.75$ has been observed in disordered films of
four different materials, grown on different substrates and measured in
different cryostats. This strongly suggests that the exponent $x\approx 0.75$
is a general property of ultrathin films of metals in the very strongly
localized regime.

The same exponent has been reported by Adkins and Astrakharchik \cite{Adkins}
in ultrathin quench-condensed films of Bi with a Ge underlayer. In that
experiment, the temperature dependence of the resistance changed to simply
activated (with $x=1$) when the Coulomb interaction was screened in the
presence of a nearby metallic gate. This behavior was ascribed to the fixed
range hopping of dipoles in screened films, but no details were given on the
origin of $x\approx 0.75$ in unscreened films. The authors suggest that the
films may be in the crossover regime between the variable range hopping and
the fixed range hopping regime. Such a crossover can occur when the optimal
hopping distance $r_{hop}$ becomes comparable with the localization length $%
\xi $.

In our experiment, $x$ stays constant over several orders of magnitude in
sheet resistance, and then drops abruptly as the sheet resistance decreases
further. If our films were merely at the crossover between $x\approx 1/2$
(or $1/3$) and $x\approx 1$, the change in $x$ would be expected to be
gradual. The observed constancy of $x$ implies that a consistent mechanism
may be governing the conduction in this regime, a mechanism different from
Mott or Efros-Shklovskii variable range hopping which is usually observed in
less resistive films.

Furthermore, it was not possible to obtain a satisfactory fit to the data
using a temperature-dependent prefactor, as in the work of Van Keuls {\it et
al}. It is interesting that these workers obtain $x\approx 0.75$ in all
magnetic fields if $m$ is taken to be zero. However, the activation energies
obtained from such fits are reported to be unacceptably small \cite{VanKeuls}%
.

There are several other mechanisms which may be relevant in a very
disordered 2D system. For example, Dai et al.\cite{Dai} observed an exponent 
$x{{}\sim }1$ in Si:B, which changed to $x{{}\sim }1/2$ when a magnetic
field was applied. They suggested that the $x{{}\sim }1$ was due to the
exchange interaction between the electron spins, which is destroyed in a
magnetic field. It must be noted that the Ag did not show any
magnetoresistance up to the highest field available, 20 kG, so the exchange
interaction is most likely not the origin of the anomalous hopping exponent
observed in these films.

If we allow a possibility that our films are granular on a very small scale
(which we cannot unequivocally rule out), than we must consider the
superlocalization mechanism of Deutscher {\it et al. }\cite{Deutscher} as a
possible candidate to explain our data. Without Coulomb interactions, this
model predicts $x\approx 0.43,$ which obviously cannot account for our
results. Including the Coulomb interactions may lead to a higher exponent,
as proposed by van der Putten {\it et al} \cite{vanderPutten}. These authors
studied the hopping conductivity of percolating carbon-black-polymer
compounds and found $x\approx 0.66,$ which is much closer to our result,
although still too low. They interpret their results as evidence of
superlocalization on a fractal network with Coulomb dominated hopping. The
activation energies were found to be independent of the electron
concentration, as predicted by Deutscher {\it et al}. In contrast, the
activation energies found in our experiment depend strongly on the film
thickness, which is closely related to the electron concentration.
Furthermore, if the Coulomb interactions were screened, one might expect the
exponent to decrease towards $0.43$, rather than to increase towards $1$, as
observed in screened Bi films by Adkins and Astrakharchik \cite{Adkins}.

Another possibility is that the anomalous exponent is a consequence of the
exponential gap in the density of states, which can arise if the electrons
interact logarithmically \cite{Shklovskii}. In that case, the hopping
exponent would be something close to, but smaller than $1$ at higher
temperatures, and then cross over smoothly to $1/2$ at low temperatures.
Forcing a fit of Eq. \ref{RT} to the data would result in an exponent which
changes continuously with temperature. In the less resistive regime where we
observe a temperature-dependent exponent, a closer inspection shows the
opposite trend: the exponent is close to $1/2$ at higher temperatures, and
increases with decreasing temperature. On the other hand, we cannot rule out
the possibility that we might observe a smooth crossover to $1/2$ in the
most resistive films if we could measure at much lower temperatures.

Finally, we consider the collective variable range hopping mechanism,
proposed in the context of vortices in disordered superconductors by Fisher 
{\it et al}. \cite{Fisher}. The mapping of this model onto\ a 2D electron
system may actually be exact, if the electron-electron interactions are
logarithmic over relevant length scales, but the authors suggest that
collective hopping may dominate over the single-particle hopping even in the
case of a conventional Coulomb interaction. The range of exponents predicted
by the collective hopping model is $2/3<x<4/5$, depending on the ratio of
electrons which participate in the process. The exponent found in our
experiment, as well as the exponents found by other groups \cite
{Adkins,VanKeuls,Gershenson,vanderPutten}, are well within that range. The
activation energies are expected to depend on the localization length, which
in turn depends on the film thickness, as shown in Fig. \ref{T0vsR}.

In conclusion, we have addressed the issue of the anomalous hopping exponent 
$x\approx 0.75$ observed in ultrathin films of metals and related 2D
systems. We argue that this hopping exponent is a general property of very
strongly disordered systems, rather than a result of an improper fit or a
signature of some sort of a crossover behavior. The usual models of hopping
conduction do not explain this result. Our data can be explained by a {\it %
collective} variable range hopping mechanism, but our work by no means
provides a proof of such a mechanism. Future experimental and theoretical
studies will be needed to shed more light on this matter.

We gratefully acknowledge useful discussions with Boris Shklovskii and
Leonid Glazman. This work was supported in part by the National Science
Foundation under Grant No. NSF/DMR-987681. 

\begin{description}
\item[{*}]  Present Address: Department of Applied Physics, Technical
University of Delft, the Netherlands.
\end{description}

\begin{description}
\item[{$^{\S }$}]  Present Address: Center for Integrated Systems, Stanford
University, Stanford, CA, USA.
\end{description}

\begin{description}
\item[{$^{\P }$}]  Present Address: Seagate Technology, Bloomington, MN, USA.
\end{description}

\begin{figure}[tbp]
\caption{Sheet resistance as a function of T$^{-0.75}$ for a series of Ag,
Bi, Pb and Pd films. Each curve represents a different thickness, and the
film thickness increases from top to bottom: a) Ag 5-6.6\AA , b) Bi 9-10\AA
, c) Pb 4.9-5.2\AA , and d) Pd 10-13\AA .}
\label{logR}
\end{figure}

\begin{figure}[tbp]
\caption{Best fits for $x$ (full circles) for increasing exponent m in the
resistance prefactor $T^{m}$ in the combination of Eqs. 1 and 2 for a 7.0$%
\AA $ thick Ag film. Chi squared (open circles) is a minimum at $m=0$. The
deviation of the fit from the data quickly becomes much larger than the
noise in $R$ as $m$ is increased. }
\label{xm}
\end{figure}

\begin{figure}[tbp]
\caption{The activation energy $T_{0}$ as a function $R_{14K}$ for different
films of all four materials: Ag (crosses), Bi (squares), Pb (diamonds) and
Pd (circles). }
\label{T0vsR}
\end{figure}

\begin{figure}[tbp]
\caption{Zabrodskii plots for a series of Ag, Bi, Pb and Pd films. Each
curve represents a different thickness, and the film thickness increases
from top to bottom: a) Ag 5-6.6\AA , b) Bi 9-10\AA , c) Pb 4.9-5.2\AA , and
d) Pd 10-13\AA . The slopes determine the exponent $x$.}
\label{zab}
\end{figure}

\begin{figure}[tbp]
\caption{The hopping exponent $x$ as a function of the sheet resistance $%
R_{14K}$ for all four materials: Ag (crosses), Bi (squares), Pb (diamonds)
and Pd (circles). }
\label{xR}
\end{figure}


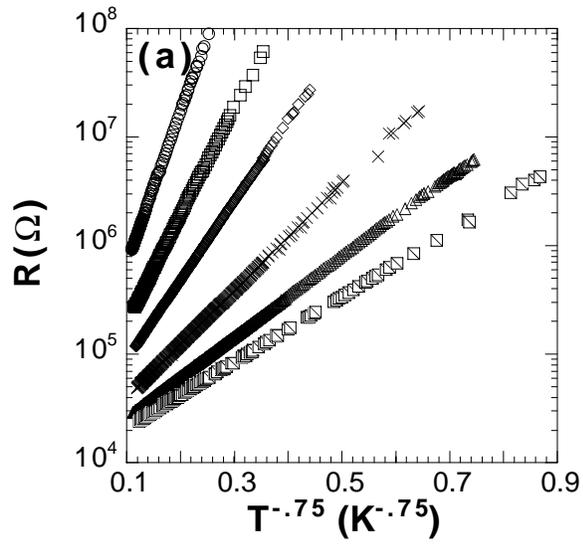
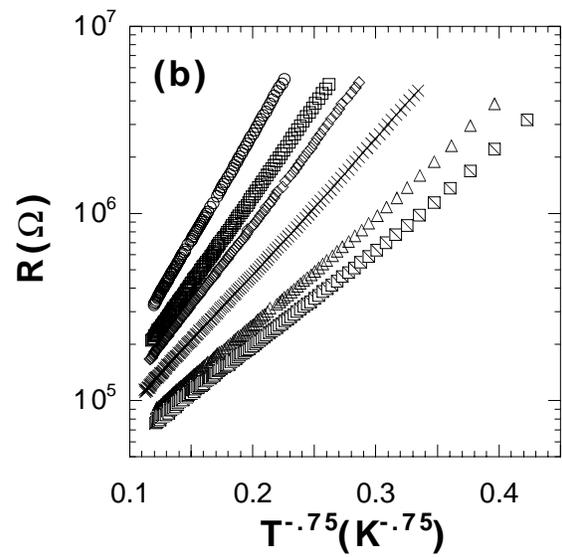
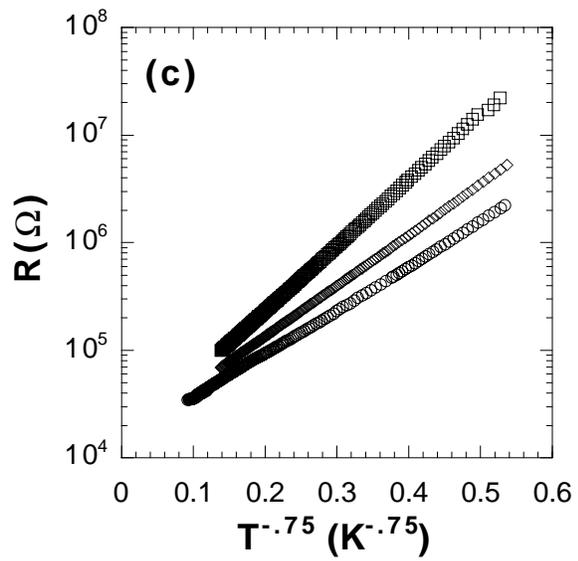
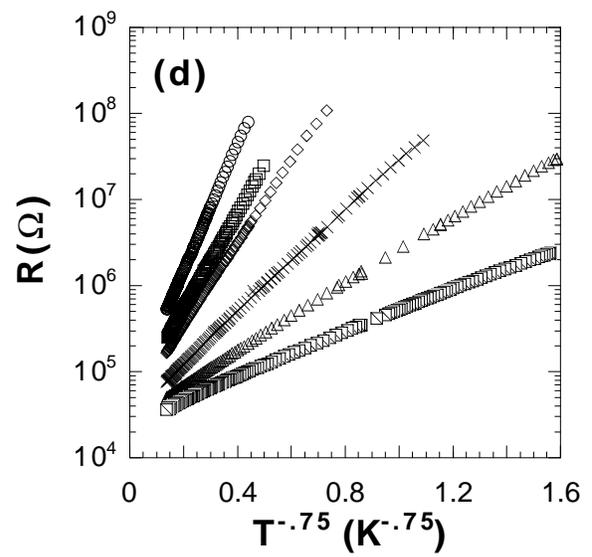
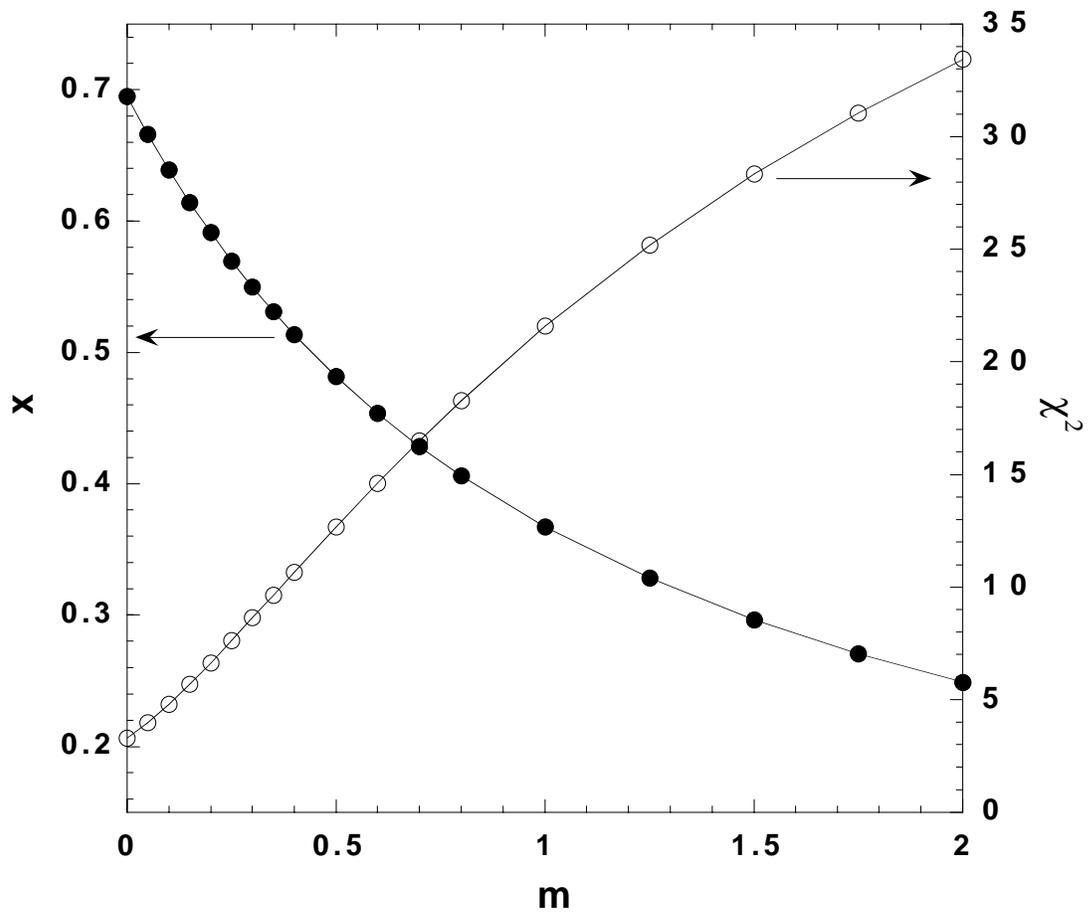
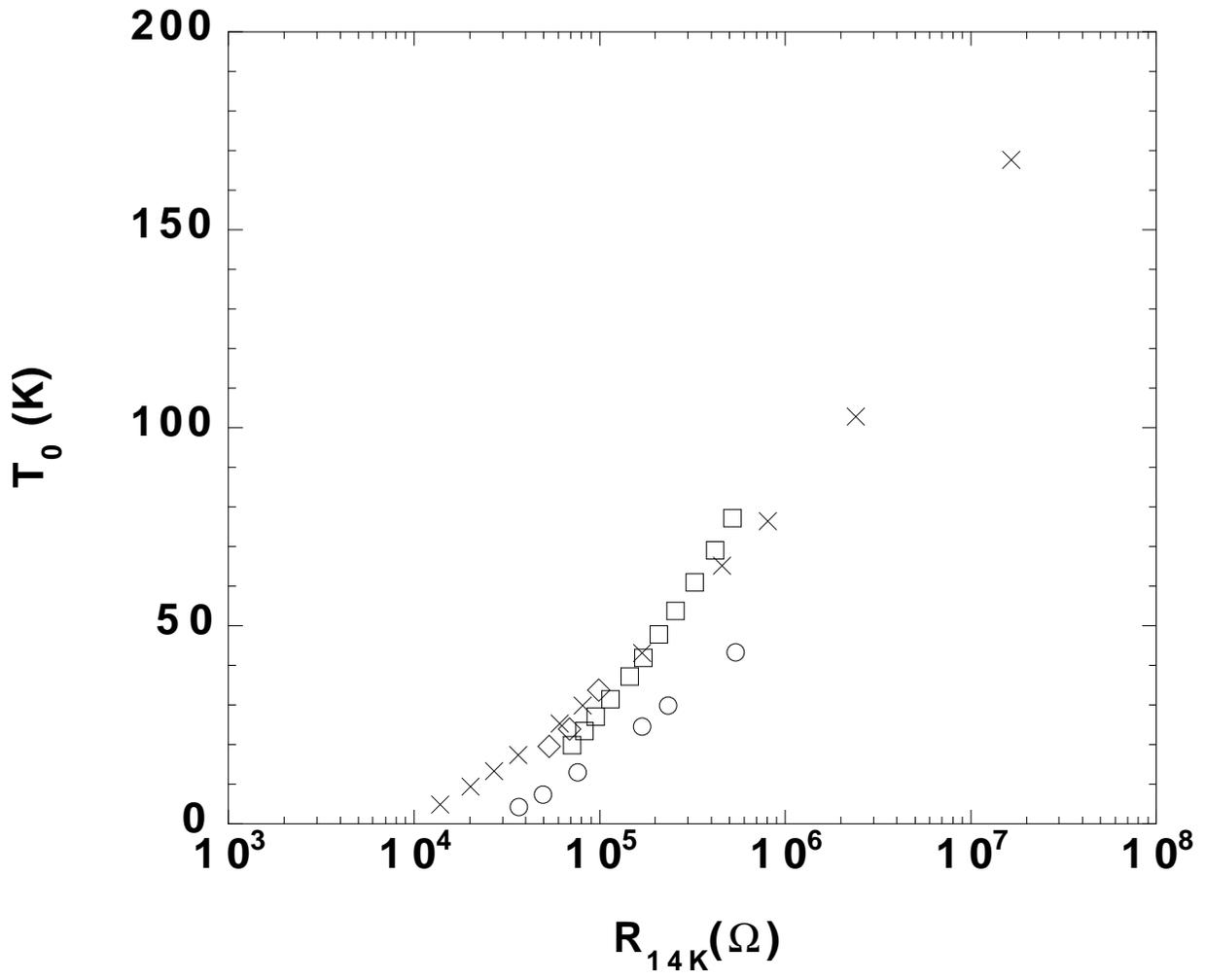
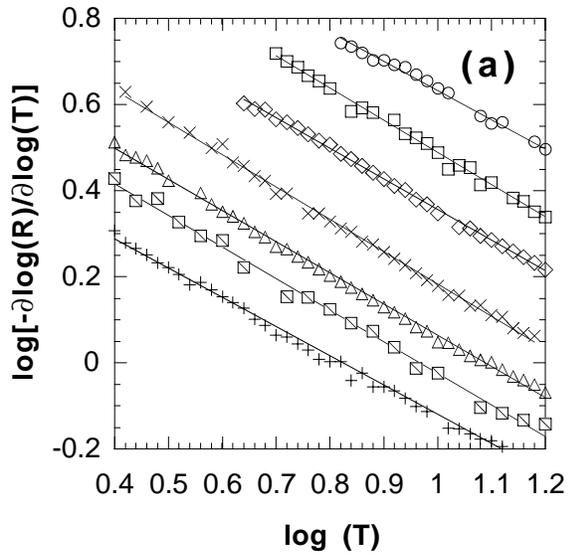
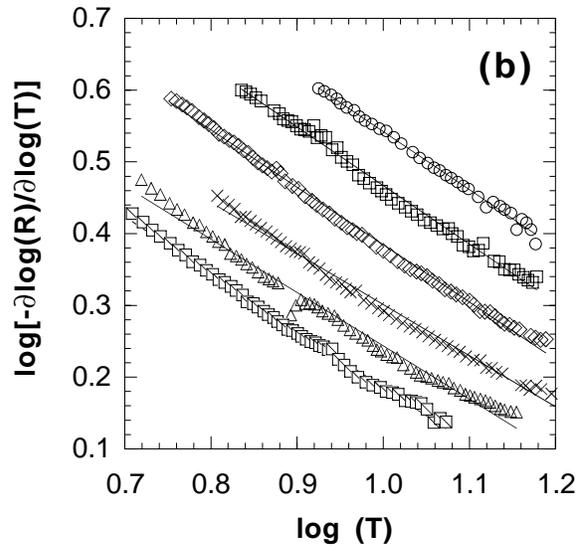
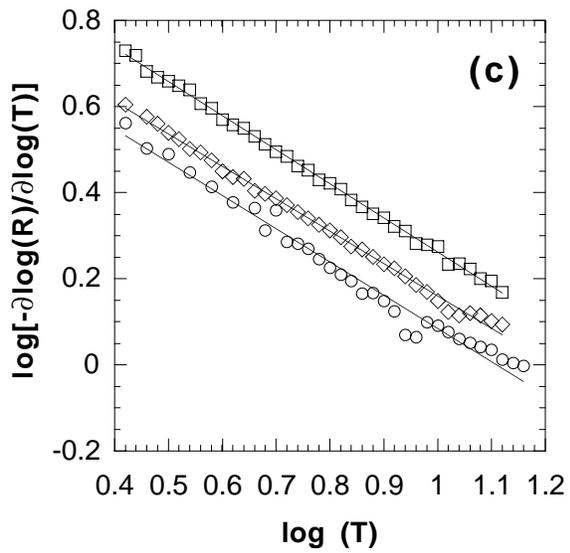
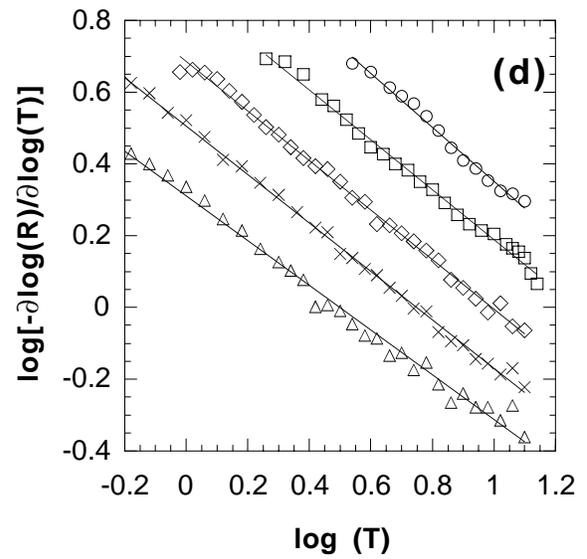
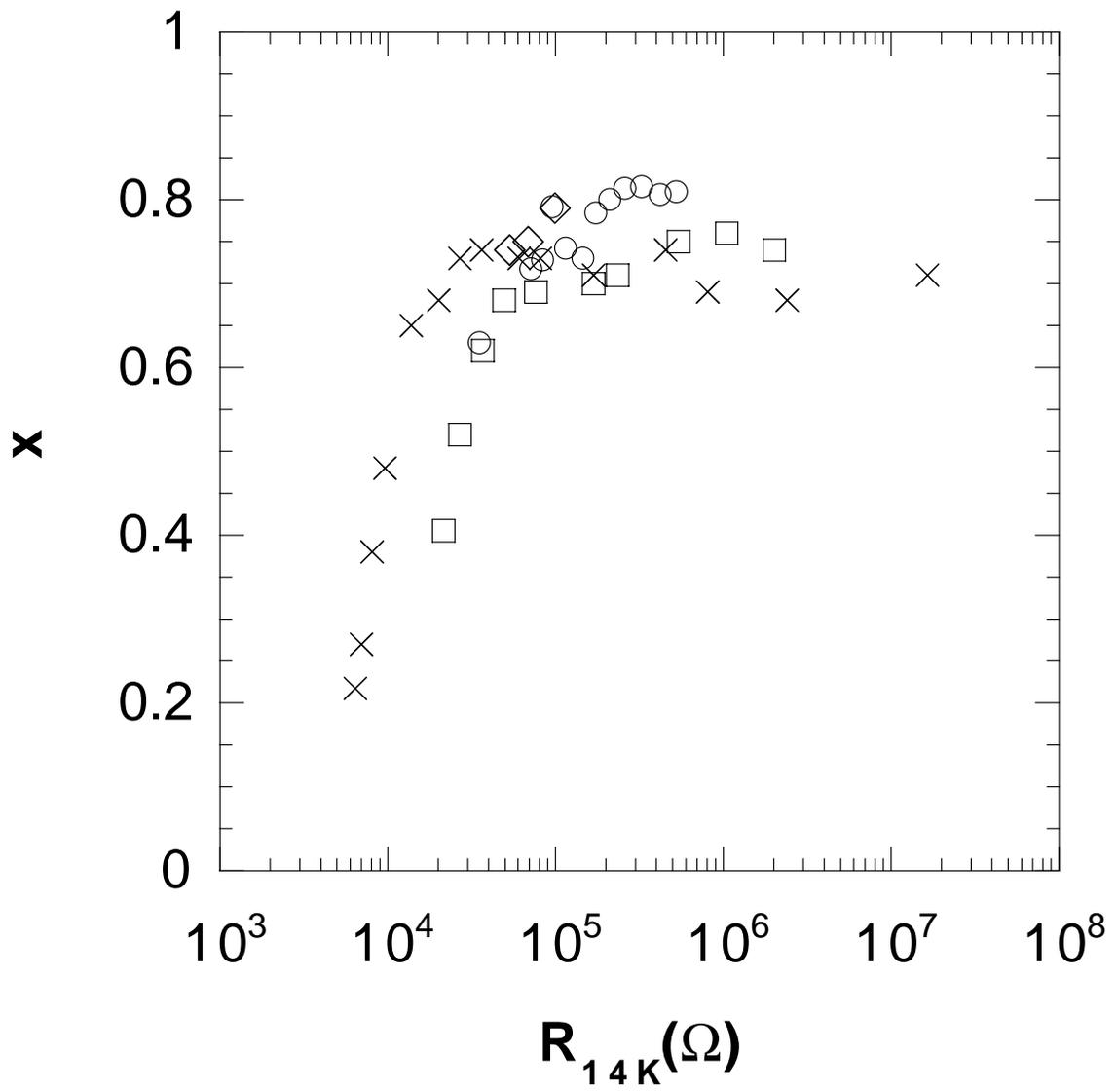
\end{document}